\documentclass[fleqn,usenatbib]{mnras}

\usepackage[T1]{fontenc}

\DeclareRobustCommand{\VAN}[3]{#2}
\let\VANthebibliography\thebibliography
\def\thebibliography{\DeclareRobustCommand{\VAN}[3]{##3}\VANthebibliography}

\usepackage{graphicx}	
\usepackage{amsmath}	
\usepackage{amssymb}	
\usepackage{epstopdf}
\usepackage{xcolor}
\usepackage{newtxtext,newtxmath}

\title[Effect of metallicity on atmospheres]{Modelling the effect of stellar metallicity on the XUV evolution of low-mass stars and its impact on exoplanet atmospheres/habitability}

\author[V. See et al.]{
Victor See$^{1,2}$\thanks{E-mail: w.c.v.see@bham.ac.uk},
Charlotte Fairman$^{3}$,
Louis Amard$^{4,5}$,
Oliver Hall$^{2}$
\\
$^{1}$School of Physics \& Astronomy, University of Birmingham, Edgbaston, Birmingham B15 2TT, UK\\
$^{2}$European Space Agency (ESA), European Space Research and Technology Centre (ESTEC), Keplerlaan 1, 2201 AZ Noordwijk, The Netherlands\\
$^{3}$HH Wills Physics Laboratory, University of Bristol, Tyndall Avenue, Bristol, BS8 1TL, UK\\
$^{4}$Department of Astronomy, University of Geneva, Chemin Pegasi 51, 1290 Versoix, Switzerland\\
$^{5}$D\'{e}partement d’Astrophysique/AIM, CEA/IRFU, CNRS/INSU, Univ.  Paris-Saclay \& Univ.  de Paris, 91191 Gif-sur-Yvette, France\\
}

\date{Accepted XXX. Received YYY; in original form ZZZ}

\pubyear{2025}

\begin{document}
\label{firstpage}
\pagerange{\pageref{firstpage}--\pageref{lastpage}}
\maketitle

\begin{abstract}
Understanding how exoplanet atmospheres evolve is a key question in the context of habitability. One key process governing this evolution is atmospheric evaporation by stellar X-ray and EUV emission (collectively, XUV). As such, the evolution of exoplanet atmospheres is closely tied to the evolution of the host star's magnetic activity. Many studies have modelled the combined evolution of exoplanet atmospheres and their host stars. However, to date, the impact of the host star's metallicity on stellar activity/exoplanet atmosphere evolution has not been explored. In this work, we investigate how stellar metallicity affects the rotation and activity evolution of solar-like stars as well as the corresponding exoplanet atmospheric evolution. We reconfirm previous results that metal-rich stars spin down more rapidly than metal-poor stars. We also find that the XUV flux that an exoplanet in the habitable zone of its host star receives is larger when the host star is more metal-rich. As such, the atmospheres of exoplanets in the habitable zones of metal-rich stars are evaporated more rapidly than exoplanets in the habitable zones of metal-poor stars. Lastly, we find that the atmospheric evolution is most sensitive to the host star metallicity when the host star has a higher mass. In the highest mass solar-stars, the metallicity can have a larger influence on the atmospheric evolution than the initial rotation period of the star.
\end{abstract}



\section{Introduction}
\label{sec:Intro}
When assessing the habitability of an exoplanet, the first consideration is usually whether the planet lies in the habitable zone (HZ) of its host star, i.e. within the range of orbital distances where it is thought surface liquid water could exist \citep{Kasting1993}. While this is a useful first consideration, many other factors can affect a planet's habitability \citep[e.g.][]{Armstrong2016}. One of the crucial questions in this context is how exoplanet atmospheres evolve over evolutionary time-scales. This is a complex question due to the numerous processes that contribute towards atmospheric evolution. In terms of atmospheric losses, one of the dominant sources is likely to be stellar magnetic activity in the form of X-ray or EUV emission (jointly referred to as XUV emission). If an exoplanet is sufficiently strongly irradiated by these types of high energy radiation, it can lead to significant atmospheric evaporation \citep{DesEtangs2010,Ehrenreich2015,Foster2022}. As such, the evolution of an exoplanet's atmosphere is closely linked to the magnetic activity evolution of its host star. The magnetic activity evolution, and specifically the XUV evolution, of low-mass stars is still an active area of research. Numerous studies have attempted to model the stellar XUV evolution and the corresponding evolution of exoplanet atmospheres \citep[e.g.][]{Claire2012,Ribas2016,Fleming2020,Pezzotti2021,Birky2021,Fujita2022}.

It is thought that the overall magnetic activity level of low-mass stars ($M_\star \lesssim 1.3M_\odot$) is determined by the Rossby number. This dimensionless parameter, defined as the stellar rotation period over the convective turnover time, ${\rm Ro}=P_{\rm rot}/\tau$, encapsulates the interaction between rotation and convection that is thought to drive magnetic field generation within low-mass stars \citep[e.g.][]{Brun2017}. Many studies have shown that magnetic activity indicators, as well as the strength of stellar magnetic fields themselves, are well parameterised by the Rossby number \citep[e.g.][]{Noyes1984,Saar1999,Pizzolato2003,Mamajek2008,Reiners2009,Vidotto2014ZDI,See2015,See2019ZB,See2019MDot,See2025,Folsom2016,Folsom2018,Stelzer2016,Newton2017,Wright2018,Kochukhov2020,Boudreaux2022,Reiners2022,Gossage2025} with stars with small Rossby numbers being more active than stars with large Rossby numbers, up to a saturation level. Therefore, it is a combination of a star's rotation and convective properties that largely determines its activity level. As such, one needs to model a star's rotation evolution first to properly predict how its activity will evolve over time. 

Over the course of the main sequence lifetime, the rotation periods of low-mass stars are known to change by several orders of magnitude. Observations of stars in open clusters of known ages show that their rotation periods increase as a function of age \citep[e.g.][]{Meibom2009,Meibom2011,Barnes2016,Rebull2016,Douglas2016,Douglas2017,Douglas2019,Curtis2020,Dungee2022}. This rotation evolution is a result of magnetised stellar winds that carry away angular momentum causing the star's rotation to slow over time. As a consequence, the activity of low-mass stars also decreases over time. Many numerical models exist that attempt to reproduce the observed rotation evolution of low-mass stars \citep[e.g.][]{Gallet2013,Gallet2015,Brown2014,Matt2015,Johnstone2015Rotation,Amard2016,Blackman2016,vanSaders2016,Gondoin2017,Ardestani2017,See2018,Garraffo2018,Amard2019,Breimann2021,Gossage2021}. By coupling a rotation model to an activity-rotation relation, it is possible to predict the overall activity output of a star over its lifetime and the corresponding impact of that activity on exoplanetary atmospheric evolution. This type of modelling has shown that both the mass of the host star and its initial rotation period affect the overall output of magnetic activity resulting in drastically different outcomes for exoplanet atmospheres \citep{Tu2015,Johnstone2015,Johnstone2021}.

\begin{figure*}
	\includegraphics[trim=5mm 10mm 5mm 20mm,width=0.9\textwidth]{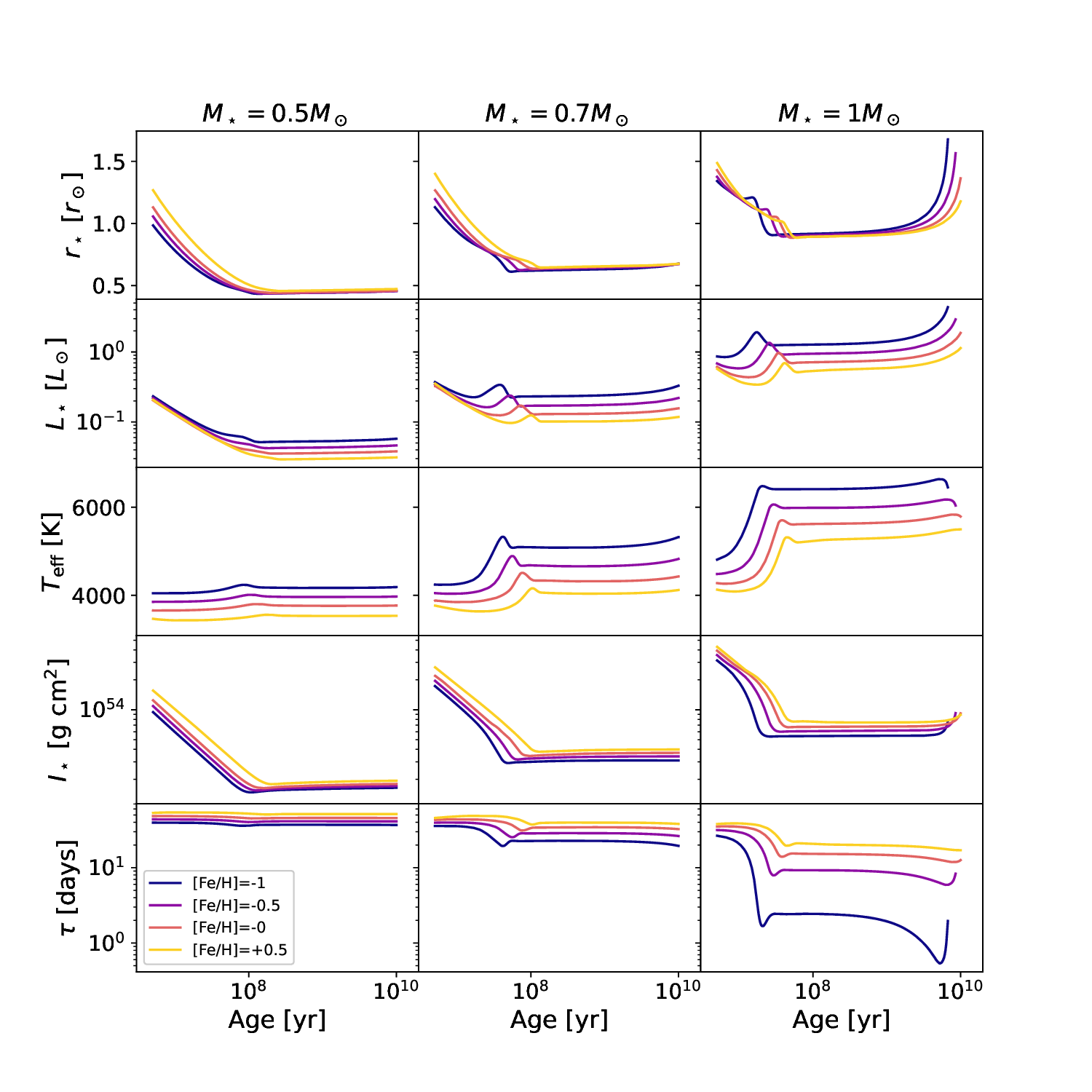}
    \caption{The evolution of the stellar radius, $r_\star$, luminosity, $L_\star$, effective temperature, $T_{\rm eff}$, moment of inertia, $I_\star$, and convective turnover time, $\tau$, of 0.5$M_\odot$ (left), 0.7$M_\odot$ (middle) and 1$M_\odot$ (right) stars. The radius, luminosity, effective temperature and moment of inertia are from the stellar structure models of \citet{Amard2019} and the turnover time is calculated using the prescription of \citet{Cranmer2011}. For each stellar mass, four different metallicities are considered: [Fe/H]=-1 (dark blue), -0.5 (purple), 0 (orange) and +0.5 (yellow).}
    \label{fig:StructModels}
\end{figure*}

While previous works have explored the impact of stellar parameters like mass or initial rotation period on the evolution of rotation, activity and exoplanet atmospheres, the role of the host star metallicity is less well understood. Recent studies have shown that various proxies of magnetic activity, like the photometric variability amplitude \citep{Karoff2018,Reinhold2020,See2021} or flaring luminosity \citep{See2023}, are stronger in metal-rich stars than in metal-poor stars. The explanation for this is that, at fixed stellar mass and rotation, more metal-rich stars should have deeper convection zones and larger convective turnover times resulting in smaller Rossby numbers and therefore stronger activity \citep{vanSaders2013,Karoff2018,Amard2019}. Although, the correlation between stellar activity and metallicity has only been demonstrated for a limited number of activity proxies to date, it is reasonable to expect that it should exist for all forms of activity since they are all ultimately driven by the stellar dynamo. One consequence of this is that metal-rich stars should lose angular momentum more rapidly than metal-poor stars because of stronger stellar winds and magnetic fields \citep{Amard2020RotEvo}. Indeed, evidence that this is the case has been found for stars in the Kepler field with metal-rich stars spinning more slowly than metal-poor stars on average \citep{Amard2020Kepler,See2024}. This then leads to an interesting question in the context of habitability. Do exoplanets around metal-rich stars receive more magnetic activity over their lifetimes than exoplanets around metal-poor stars? On the one hand, more metal-rich stars should be more magnetically active, all else being equal. On the other hand, more metal-rich stars spin down more rapidly and slow rotation is associated with weaker activity. 

\begin{figure*}
	\includegraphics[trim=0mm 10mm 0mm 0mm,width=\textwidth]{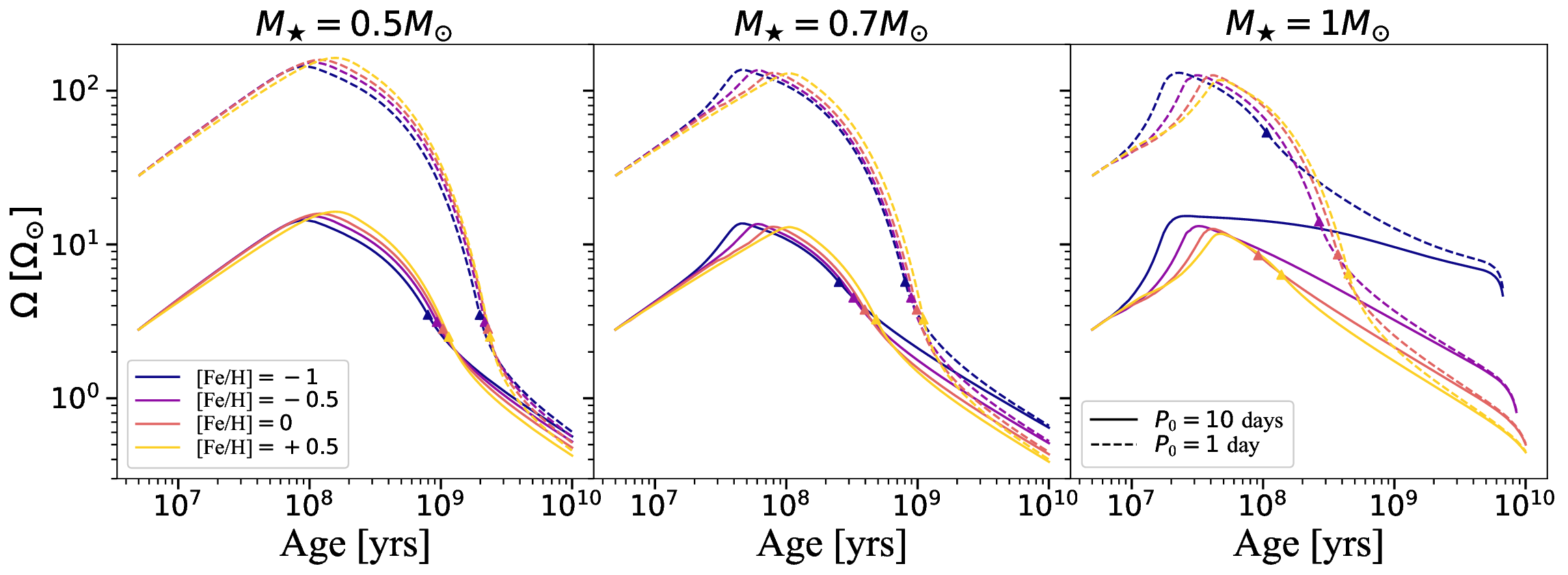}
    \caption{The modelled angular velocity evolution of stars of different masses, initial rotation periods and metallicities. The masses considered are 0.5$M_\odot$ (left panel), 0.7$M_\odot$ (middle panel) and 1$M_\odot$ (right panel). The initial periods considered are 1 day (dashed lines) and 10 days (solid lines). The metallicities considered are -1 dex (dark blue lines), -0.5 dex (purple lines), 0 dex (orange lines) and +0.5 dex (yellow lines). The point at which the star transitions from the saturated to unsaturated regime (at $\rm Ro_{\rm crit} = 0.2$) is indicated by a triangle along each track.}
    \label{fig:RotEvo}
\end{figure*}

In this work, we explore the role of metallicity on the evolution of rotation, activity and exoplanet atmospheres using parametric evolution models. In particular, we focus on the evolution of the stellar XUV flux and the rate at which it evaporates exoplanet atmospheres. The rest of the paper is structured as follows. In section \ref{sec:StructModels}, we show how metallicity affects the physical properties of low-mass stars. We present our rotation evolution model in section \ref{sec:RotEvo}. In section \ref{sec:ActEvo}, we couple our rotation evolution model to an activity-rotation relation to model the XUV evolution of low-mass stars. In section \ref{sec:Exo}, we use the XUV evolution to model the rate at which exoplanet atmospheres are evaporated. We discuss various caveats and implications of our results in section \ref{sec:Discussion} and present our conclusions in section \ref{Sec:Conclusions}.

\section{Stellar structure models}
\label{sec:StructModels}
The aim of this study is to understand the impact that stellar metallicity has on the rotation and activity of low-mass stars and the atmospheres of any exoplanets that may orbit them. Ultimately, in our models, the metallicity affects the stellar and planetary evolution via its impact on the physical structure of the star. It will, therefore, be useful to explicitly look at how the physical properties of low-mass stars vary as a function of metallicity. In this work, we use the stellar structure models of \citet{Amard2019}. Specifically, we will consider models at three stellar masses (0.5$M_\odot$, 0.7$M_\odot$, 1.0$M_\odot$) and with four metallicities ([Fe/H] = -1.0, -0.5, 0.0, +0.5) for a total combination of 12 different structure models. We choose not to study stars less massive than 0.5$M_\odot$ since the rotation evolution of these stars is still not well understood. In fig. \ref{fig:StructModels}, we show the evolution of the stellar parameters from the structure models that will be relevant throughout this work as a function of mass and metallicity, i.e. the stellar radius, luminosity, effective temperature and moment of inertia. Additionally, we also show the convective turnover time. This is calculated using the prescription of \citet{Cranmer2011} and is given by 

\begin{equation}
    \tau = 314.24 \exp\left[-\left(\frac{T_{\rm eff}}{1952.5{\rm K}}\right) - \left(\frac{T_{\rm eff}}{6250 {\rm K}}\right)^{18}\right]+0.002,
    \label{eq:Turnover}
\end{equation}
where $T_{\rm eff}$ is the effective temperature (we discuss this choice of turnover time prescription in greater detail in section \ref{subsec:TurnoverTimes}). Looking at the metallicity dependence, we see that metal-rich stars are generally larger, less luminous, cooler, have larger moments of inertia and longer turnover times than metal-poor stars for a given stellar mass and age.

\section{Rotation evolution}
\label{sec:RotEvo}
\subsection{Model description}
\label{subsec:RotEvoModel}
To model the rotation evolution of a star, we numerically solve the angular momentum equation which is given by

\begin{equation}
    \frac{d \Omega_\star}{dt} = \frac{T}{I_\star} - \frac{\Omega_\star}{I_\star} \frac{dI_\star}{dt},
    \label{eq:AngMomEq}
\end{equation}
where $\Omega_\star$ is the stellar angular velocity, $T$ is the torque or angular momentum loss rate associated with the stellar wind, $I_\star$ is the star's momentum of inertia and $t$ is time. While many different prescriptions exist to estimate the stellar torque from the properties of the star \citep[e.g.][]{Matt2012,vanSaders2013,Reville2015,Garraffo2016,Pantolmos2017,Finley2018}, we use the prescription of \citet{Matt2015} which is given by

\begin{equation}
\begin{split}
    T &= -T_0 \left(\frac{\tau_\star}{\tau_\odot}\right)^2 \left(\frac{\Omega_\star}{\Omega_\odot}\right)^3 \hspace{5mm} &{\rm unsaturated}\\
    T &= -100T_0 \left(\frac{\Omega_\star}{\Omega_\odot}\right) &{\rm saturated}, 
\end{split}    
\label{eq:MattTorque}
\end{equation}
where $\tau$ is the convective turnover time. The normalising term, $T_0$, is given by

\begin{equation}
    T_0 = 6.3 \times 10^{30} \left(\frac{r_\star}{r_\odot}\right)^{3.1} \left(\frac{M_\star}{M_\odot}\right)^{0.5},
    \label{eq:TorqueNorm}
\end{equation}
where $r_\star$ is the stellar radius and $M_\star$ is the stellar mass. The critical Rossby number separating the unsaturated and saturated regimes in equation (\ref{eq:MattTorque}) is chosen to be $\rm Ro_{\rm crit} = 0.1Ro_\odot = 0.2$. The stellar parameters needed to solve equations (\ref{eq:AngMomEq})-(\ref{eq:TorqueNorm}) are taken from the stellar structure models of \citet{Amard2019} as shown in section \ref{sec:StructModels}. Similarly to \citet{Matt2015}, solid body rotation is assumed.

\subsection{Model results}
\label{subsec:RotEvoResults}
Fig. \ref{fig:RotEvo} shows the results of our rotation evolution modelling. Each model has an initial rotation period of either $P_{\rm 0}=$1 day or $P_{\rm 0}=$10 days at an age of 5 Myr to represent the spread of initial rotation periods that stars are known to possess at early ages \citep[e.g.][]{Irwin2008,Rebull2018}. The effect of metallicity on the rotation evolution of low-mass stars has already been explored by \citet{Amard2020RotEvo}. Although there are some small differences between our rotation model and the ones presented by these authors, our results are in broad agreement with theirs. Here, we briefly discuss the main features of fig. \ref{fig:RotEvo} and refer the interested reader to \citet{Amard2020RotEvo} for a more complete discussion of how metallicity affects the rotation evolution of low-mass stars.

During the pre-main sequence phase, stars spin up in order to conserve angular momentum as they contract, with metal-rich stars taking longer to finish contracting than metal-poor stars. As such, we see that more metal-rich stars take longer to reach their maximum rotation rate at early ages than metal-poor stars for a given mass and initial rotation period in fig. \ref{fig:RotEvo}. Once on the main sequence, stars spin down due to their magnetised winds. We see that metal-rich stars of a given mass spin down more rapidly than metal-poor stars and reach slower rotation at late ages. This is because metal-rich stars have longer turnover times, which can be seen in fig. \ref{fig:StructModels}, resulting in a larger spin-down torque in the unsaturated regime in equation (\ref{eq:MattTorque}). Figure \ref{fig:RotEvo} also demonstrates that stars of a given mass and a given metallicity converge onto a single rotation period vs age sequence regardless of their initial rotation period with the location of this sequence being determined by the metallicity.

Lastly, we discuss the effect of mass in fig. \ref{fig:RotEvo}. The effect of stellar mass on rotation evolution has previously been discussed in detail by \citet{Matt2015} and our results broadly reflect the conclusions drawn by these authors. The rotation evolution of all three stellar masses shown in fig. \ref{fig:RotEvo} display the same qualitative behaviour. The only differences are the times at which certain features occur, e.g. the transition from the saturated to unsaturated regime. One interesting point that is worth noting is that the spread in rotation rates at late ages due to the different metallicities is larger in higher mass stars. The reason for this is that higher mass stars have thinner convective envelopes than lower mass stars. Therefore, a change in metallicity of a given amount results in a larger fractional change to the convective properties of higher mass stars which can be seen in the convective turnover times shown in fig. \ref{fig:StructModels} \citep[see also][]{Amard2020RotEvo}. As such, the fractional change in the torque of higher mass stars is also correspondingly larger. We previously suggested that a similar trend seemed to occur when looking at magnetic activity \citet{See2021}. In that work, we noted that the photometric variability amplitude, a proxy for activity, appeared to be more sensitive to changes in metallicity in higher mass stars.

\section{Activity evolution}
\label{sec:ActEvo}
In this section, we model the XUV evolution of low-mass stars of different metallicities. We first re-analyse the data from \citet{Wright2011} in section \ref{subsec:DetermineXUV} to obtain an X-ray - Rossby number relation that is consistent with our rotation evolution model. We then estimate the stellar EUV emission from the X-ray emission. In section \ref{subsec:XUVLumEvoResults}, we determine the evolution of the XUV luminosity by coupling the activity-rotation relations to our rotation evolution model. Finally, in section \ref{subsec:XUVFluxEvoResults}, we determine the evolution of the XUV flux received by a planet within the HZ.

\subsection{Determining the XUV flux}
\label{subsec:DetermineXUV}
In this section, we outline the method we use to estimate a star's XUV emission. To begin, we derive an X-ray vs Rossby number relation using the data set of \citet{Wright2011}. These authors compiled a sample of $\sim$800 stars with X-ray and rotation period measurements. They also empirically determined convective turnover times for their stars. However, we re-derive convective turnover times for this sample of stars using equation (\ref{eq:Turnover}) and the effective temperatures compiled by \citet{Wright2011}. We do this so that the turnover times we use to construct our X-ray vs Rossby number relation are consistent with the ones used in our rotation evolution model in section \ref{sec:RotEvo}. Since equation (\ref{eq:Turnover}) only applies for stars with $T_{\rm eff}\gtrsim 3300 {\rm K}$ \citep[see][]{Cranmer2011}, we exclude any stars cooler than this in the following analysis. We also exclude any stars hotter than $T_{\rm eff} = 6400 {\rm K}$ as this is approximately where the Kraft break occurs (the Kraft break is the temperature at which the convective envelopes of solar-like stars become vanishingly thin resulting in a breakdown of the activity-rotation relations for hotter stars).

\begin{figure}
	\includegraphics[trim=5mm 10mm 5mm 10mm,width=\columnwidth]{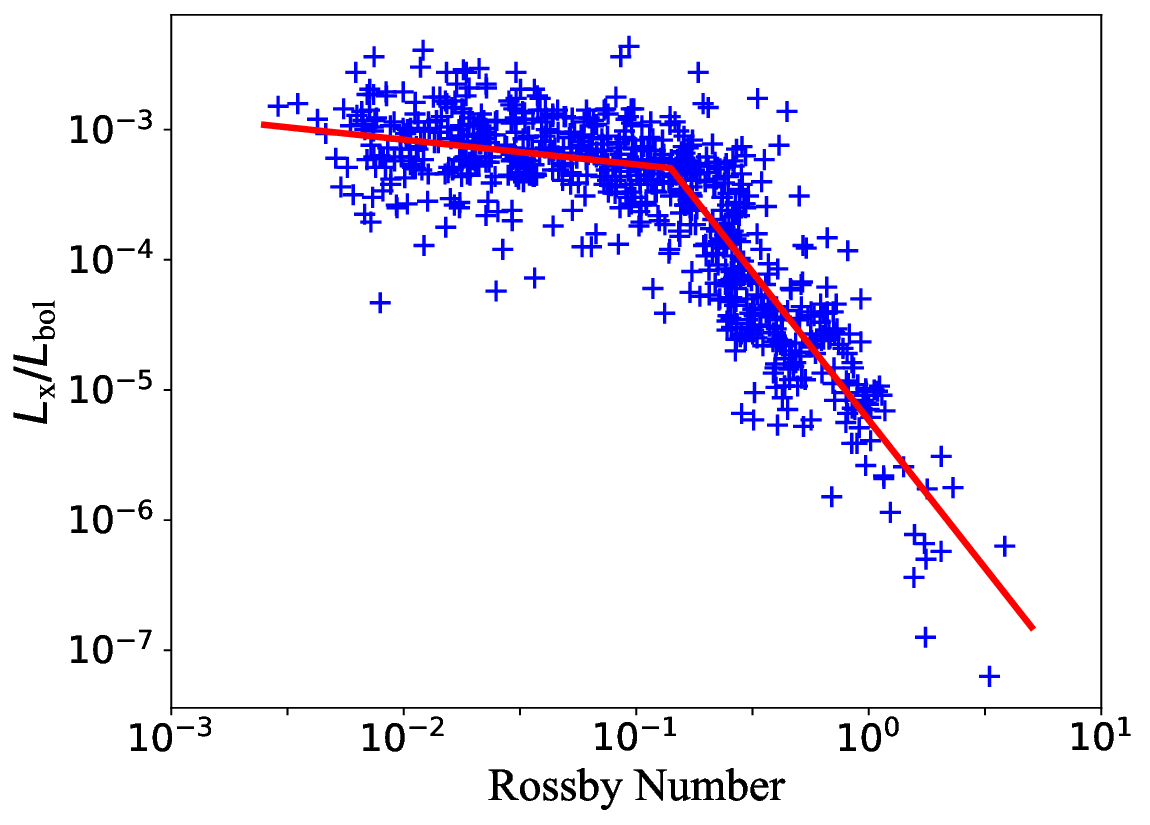}
    \caption{X-ray to bolometric luminosity ratio as a function of Rossby number. The data is taken from \citet{Wright2011} with the convective turnover times, and therefore Rossby numbers, recalcuated using the prescription of \citet{Cranmer2011}. Our best fit line takes the form of a broken power law (equation (\ref{eq:XrayActRotRel})) and is shown in red.}
    \label{fig:ActRotRel}
\end{figure}

Figure \ref{fig:ActRotRel} shows the X-ray luminosity over the bolometric luminosity, $R_{\rm X} = L_{\rm X}/L_{\rm bol}$, against Rossby number using our newly re-derived turnover times. We fit a broken power law to this data of the form

\begin{equation}
\begin{split}
    \log (L_{\rm X}/L_{\rm bol}) &= \beta_{\rm sat} \log {\rm Ro} + C_{\rm sat} \hspace{10mm} {\rm Ro\leq Ro_{sat}} \\
    \log (L_{\rm X}/L_{\rm bol}) &= \beta_{\rm unsat} \log {\rm Ro} + C_{\rm unsat} \hspace{5mm} {\rm Ro> Ro_{sat}},
\end{split}
\label{eq:XrayActRotRel}
\end{equation}
where the values of the fit parameters are $\beta_{\rm sat}=-0.19\pm 0.04$, $C_{\rm sat}=-3.46\pm 0.07$, $\beta_{\rm unsat}=-2.27\pm 0.07$, $C_{\rm unsat}=-5.23\pm 0.04$. With these best fit parameters, the critical Rossby number that separates the saturated and unsaturated regimes occurs at $\rm Ro_{sat}=0.14$. This fit is shown in fig. \ref{fig:ActRotRel} with a red line. When fitting activity-rotation relations, it is common to force the fit in the saturated regime to be flat, i.e. $\beta_{\rm sat}=0$. However, we have not enforced such a requirement resulting in a shallow, but non-zero, slope in the saturated regime of our fit. A number of other authors have also noted this shallow slope in the X-ray saturated regime \citep{Reiners2014,Magaudda2020,Johnstone2021}.

\begin{figure*}
	\includegraphics[trim=0mm 0mm 0mm 0mm, clip, width=0.95\textwidth]{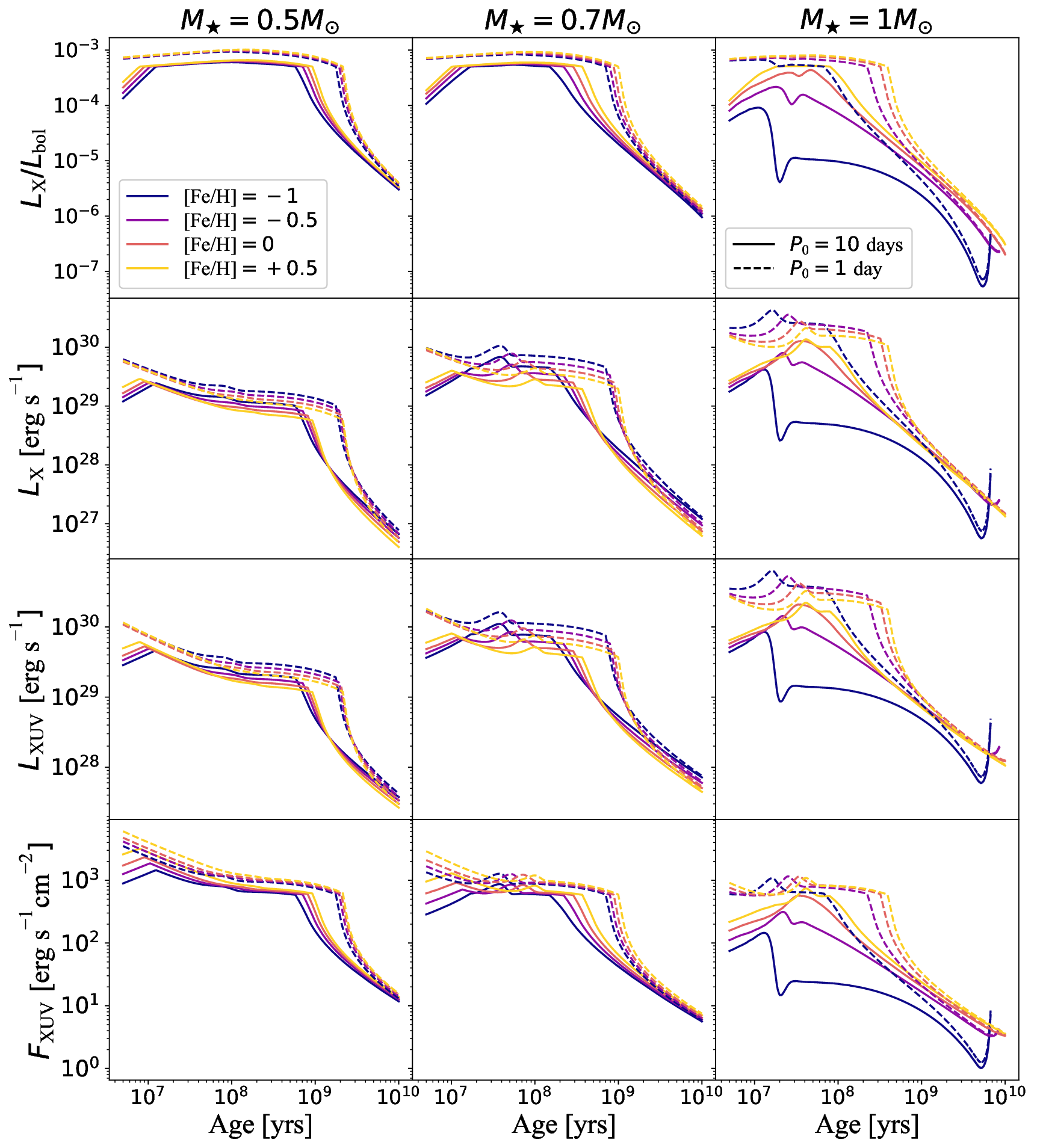}
    \caption{The modelled evolution of the X-ray to bolometric luminosity ratio, $R_{\rm X}=L_{\rm X}/L_{\rm bol}$ (top row), X-ray luminosity, $L_{\rm X}$ (second row), XUV luminosity, $L_{\rm XUV}$ (third row), and XUV flux at the midpoint of the habitable zone for an Earth mass planet, $F_{\rm XUV}(r_{\rm HZ,mid})$ (bottom row), for 0.5$M_\odot$ (left column), 0.7$M_\odot$ (middle column) and 1$M_\odot$ (right column) stars. A range of metallicities and initial rotation periods are considered and are indicated by the same colour and line style scheme as fig. \ref{fig:RotEvo}.}
    \label{fig:XUVLumEvo}
\end{figure*}

Determining the EUV emission of low-mass stars is more difficult than determining their X-ray emission because fewer observational constraints exist in the EUV. Observations are only possible for the closest stars because of interstellar absorption. Rather than estimating the EUV emission of low-mass stars from their Rossby number in a similar way to the X-ray emission, it is typical to estimate the stellar EUV emission from their X-ray emission using empirical relations \citep[e.g.][]{Sanz-Forcada2011,Chadney2015,King2021}. In this work, we use the following relation from \citet{Johnstone2021} that links the surface EUV flux of a star to its surface X-ray flux,

\begin{equation}
\begin{split}
    \log F_{\rm EUV,1}(r_\star) &= 2.04+0.681 \log F_{\rm X}(r_\star) \\
    \log F_{\rm EUV,2}(r_\star) &= -0.341 + 0.92\log F_{\rm EUV,1}(r_\star),
\end{split}
\label{eq:FEUVvsFX}
\end{equation}
where $F_{\rm EUV,1}(r_\star)$ is the surface EUV flux in the 10-36 nm range, $F_{\rm EUV,2}(r_\star)$ is the surface EUV flux in the 36-92 nm range and the total EUV surface flux is given by their sum, i.e. $F_{\rm EUV}(r_\star) = F_{\rm EUV,1}(r_\star) + F_{\rm EUV,2}(r_\star)$. These authors constructed this relation using a small sample of nearby F, G, K and M stars with EUV and X-ray measurements as well as observations of the Sun.

In section \ref{sec:Exo}, it is the XUV flux at the orbit of the exoplanet, $F_{\rm XUV}(r_{\rm orb})$, that we will require to calculate the rate at which the exoplanet atmosphere is being evaporated. The process for calculating this is as follows. First, we calculate $L_{\rm X}$ from the star's Rossby number using equation (\ref{eq:XrayActRotRel}). This step requires the star's rotation period, convective turnover time and luminosity. The rotation period comes from our rotation evolution modeling in section \ref{subsec:RotEvoResults}, the turnover time comes from equation (\ref{eq:Turnover}) and the luminosities come from the models of \citet{Amard2019}. Then we convert $L_{\rm X}$ to an X-ray flux at the stellar surface, $F_{\rm X}(r_\star)=L_{\rm X}/4\pi r_\star^2$. The stellar radius needed for this calculation is also taken from the stellar structure models of \citet{Amard2019}. Next, we calculate the surface EUV flux, $F_{\rm EUV}(r_\star)=F_{\rm EUV,1}(r_\star)+F_{\rm EUV,2}(r_\star)$, from the surface X-ray flux using equation (\ref{eq:FEUVvsFX}). Finally, we sum the surface X-ray and EUV fluxes, $F_{\rm XUV}(r_\star)=F_{\rm X}(r_\star) + F_{\rm EUV}(r_\star)$, and scale it by the square of the orbital distance of the planet to determine the XUV flux at the planet, $F_{\rm XUV}(r_{\rm orb}) = F_{\rm XUV}(r_\star)(r_\star / r_{\rm orb})^2$.

\subsection{XUV luminosity evolution}
\label{subsec:XUVLumEvoResults}
The activity evolution of a star can be modelled using the method outlined in section \ref{subsec:DetermineXUV} in conjunction with the rotation evolution of the star shown in section \ref{subsec:RotEvoModel}. Figure \ref{fig:XUVLumEvo} shows the results of this process. The top row of fig. \ref{fig:XUVLumEvo} shows the modelled X-ray to bolometric luminosity evolution, $R_{\rm X}=L_{\rm X}/L_{\rm bol}$. The $R_{\rm X}$ behaviour is easiest to understand if we initially focus on 0.5$M_\odot$ stars. For every initial rotation period and metallicity we consider, 0.5$M_\odot$ stars spend the majority of their early to middle life in the saturated regime of X-ray activity. As such, for a given star, $R_{\rm X}$ remains relatively constant over this initial evolutionary phase, with the exception of the first few Myr for the stars with $P_0$=10 days when they are rotating slow enough to be in the X-ray unsaturated regime. It is worth noting that the $P_0$=1 day stars have larger $R_{\rm X}$ values over this initial saturated phase than the stars with $P_0$=10 days. This is due to the fact that we did not enforce a flat dependence for $R_{\rm X}$ in the saturated regime of our fit of equation (\ref{eq:XrayActRotRel}). Once in the unsaturated regime, the $R_{\rm X}$ decays rapidly for all stars with the transition from the saturated to unsaturated regime occurring earlier for stars with $P_0$=10 days ($\sim$800 Myr) than stars with $P_0$=1 day ($\sim$2 Gyr). When considering the metallicity dependence, we see that metal-rich stars always have larger $R_{\rm X}$ values than metal-poor stars for a given initial rotation period even at late ages when metal-rich stars have spun down to longer rotation periods than their metal-poor counterparts. The $R_{\rm X}$ behaviour for 0.7$M_\odot$ stars is qualitatively similar to that of 0.5$M_\odot$ stars. The main differences are that 0.7$M_\odot$ stars leave the saturated regime earlier and reach smaller $R_{\rm X}$ values at late ages than 0.5$M_\odot$ stars. The $R_{\rm X}$ evolution of 1$M_\odot$ stars is also qualitatively similar to that of 0.5$M_\odot$ and 0.7$M_\odot$ stars with one notable difference. During the pre-main sequence, all 0.5$M_\odot$ and 0.7$M_\odot$ stars spin up enough to enter the X-ray saturated regime. However, the three most metal poor 1$M_\odot$ stars with $P_0$=10 days never rotate rapidly enough to be in the X-ray saturated regime. As such, these stars show much larger $R_{\rm X}$ variations during the early phases of evolution than stars that spend time in the saturated regime. The last point worth noting is that the spread in $R_{\rm X}$ values that is associated with the spread in metallicities, for stars of a given mass and initial rotation period, is larger in higher mass stars. As discussed in section \ref{subsec:RotEvoResults}, this is because higher mass stars have thinner convective zones and a given change in metallicity results in a larger fractional change in their convective properties resulting in a larger change of their activity. 

The second row of fig. \ref{fig:XUVLumEvo} shows the modelled X-ray luminosity evolution of our stars. This is calculated by taking the $R_{\rm X}$ evolution from the top row of fig. \ref{fig:XUVLumEvo} and multiplying it by the bolometric luminosities obtained from the stellar structure models of \citet{Amard2019} from section \ref{sec:StructModels}. The $L_{\rm X}$ evolution is therefore qualitatively similar to the $R_{\rm X}$ evolution with a few important differences. The first is related to the effect of the stellar mass. All else being equal, higher mass stars have larger bolometric luminosities. Therefore, for a given $R_{\rm X}$, higher mass stars will have larger $L_{\rm X}$. As such, our modelling shows that higher mass stars generally have larger X-ray luminosities than lower mass stars, at a given age, which is consistent with the results of \citet{Johnstone2021}. The second difference between the $R_{\rm X}$ and $L_{\rm X}$ evolution is related to the effect of metallicity. Metal-poor stars have larger bolometric luminosities than metal-rich stars for a given stellar mass. When converting from $R_{\rm X}$ to $L_{\rm X}$, this can result in metal-poor stars having larger $L_{\rm X}$ values than metal-rich stars at certain phases of their evolution for a given stellar mass and initial rotation period. This is notable because the same is never true of $R_{\rm X}$, i.e. $R_{\rm X}$ is always larger in metal-rich stars than in metal-poor stars for a given stellar mass, initial rotation period and age.

The third row of fig. \ref{fig:XUVLumEvo} shows the evolution of the XUV luminosity. When we outlined the steps for calculating the XUV flux at the planetary orbit, $F_{\rm XUV}(r_{\rm orb})$, in section \ref{subsec:DetermineXUV}, we did not explicitly calculate the XUV luminosity. Instead, $F_{\rm XUV}(r_{\rm orb})$ is calculated from the surface XUV flux, $F_{\rm XUV}(r_\star)$. However, in fig. \ref{fig:XUVLumEvo}, we choose to show the XUV luminosity rather than $F_{\rm XUV}(r_\star)$ since it is easier to compare to the X-ray luminosity shown in the second row of fig. \ref{fig:XUVLumEvo}. Since there is a simple scaling relation between the EUV and X-ray via equation (\ref{eq:FEUVvsFX}), the XUV luminosity evolution is qualitatively the same as the X-ray luminosity evolution shown in the second row of fig. \ref{fig:XUVLumEvo}.

\begin{figure}
	\includegraphics[trim=0mm 10mm 0mm 0mm,width=\columnwidth]{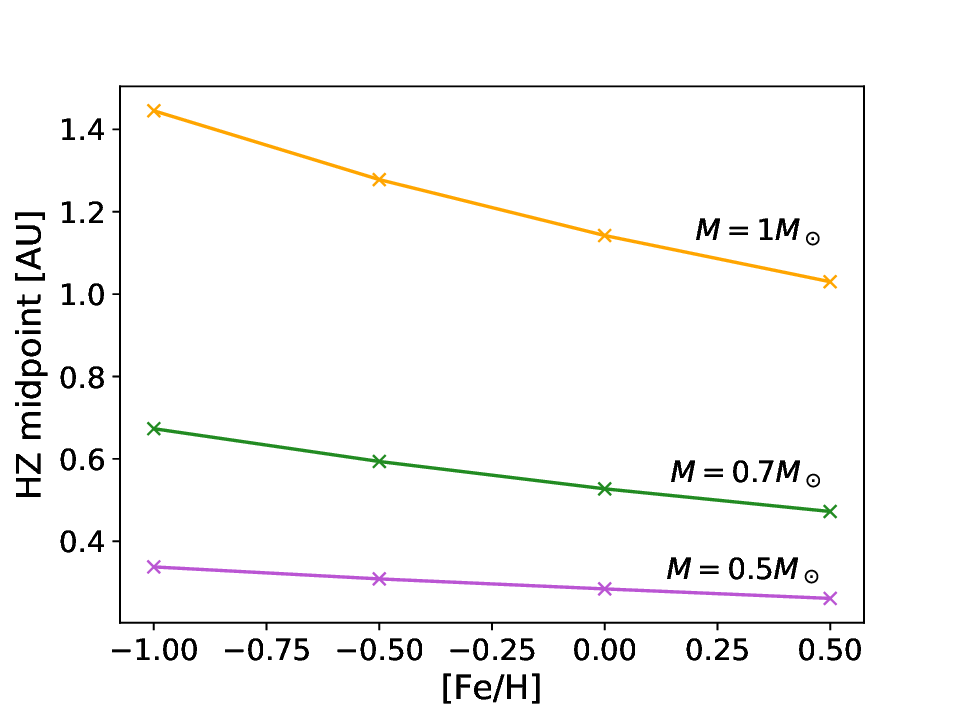}
    \caption{The distance to the center of the habitable zone as a function of metallicity for an Earth mass planet around 0.5$M_\odot$ (purple), 0.7$M_\odot$ (green) and 1$M_\odot$ (orange) stars. The distances are calculated using equations (\ref{eq:HZDistance}) and (\ref{eq:EffFlux}) with the effective temperatures and luminosities taken from the structure models of \citet{Amard2019} at an age of 500 Myr.}
    \label{fig:HZDist}
\end{figure}

\begin{table}
	\centering
	\caption{Coefficients given by \citet{Kopparapu2014} to calculate the inner and outer habitable zone boundaries in equation (\ref{eq:HZDistance}). For the inner habitable zone, values are given for a 0.1, 1.0 and 5.0 Earth Mass planet.}
	\label{tab:HZParams}
	\begin{tabular}{lccccc}
		\hline
		& $S_{\rm eff,\odot}$ & $a\ (10^{-4})$ & $b\ (10^{-8})$ & $c\ (10^{-12})$ & $d\ (10^{-15})$\\
		\hline
        Inner (0.1$M_\oplus$) & 0.99 & 1.209 & 1.404 & -7.418 & -1.713 \\
		Inner (1$M_\oplus$) & 1.107 & 1.332 & 1.58 & -8.308 & -1.931 \\
        Inner (5$M_\oplus$) & 1.188 & 1.433 & 1.707 & -8.968 & -2.084 \\
		Outer & 0.356 & 0.6171 & 0.1698 & -3.198 & -0.5575 \\
		\hline
	\end{tabular}
\end{table}

\subsection{XUV flux at the habitable zone location}
\label{subsec:XUVFluxEvoResults}
In order to calculate the XUV flux at the exoplanet, we must first pick an orbital radius for the planet. Since our aim is to investigate exoplanetary habitability, we will assume that the exoplanet lies at the center of the habitable zone of the star, i.e. $r_{\rm orb}=r_{\rm HZ,mid}$. In this work, we use the HZ prescription of \citet{Kopparapu2014}. The distance to the HZ boundaries are given by

\begin{equation}
    r_{\rm HZ} = \left(\frac{L_{\rm bol}/L_\odot}{S_{\rm eff}} \right)^{0.5} {\rm AU}.
    \label{eq:HZDistance}
\end{equation}
Here, the effective stellar flux, $S_{\rm eff}$ is given by

\begin{equation}
    S_{\rm eff} = S_{\rm eff,\odot} + a T_\star + b T_\star^2 + c T_\star^3 + d T_\star^4,
    \label{eq:EffFlux}
\end{equation}
where $T_\star = T_{\rm eff} - 5780 {\rm K}$ and the values of $S_{\rm eff,\odot}$, $a$, $b$, $c$ and $d$ are given in table \ref{tab:HZParams} for the the inner and outer edges of the HZ for a range of planetary masses. The luminosities and effective temperatures required for these calculations are taken from the structure models of \citet{Amard2019}. Since the luminosity and temperature of a star change over the stellar lifetime, the HZ location also changes \citep{Gallet2017}. In this work, we calculate the HZ midpoint at an age of 500 Myr, i.e. an age when all the stars we consider are on the main sequence. We then make the simplifying assumption that the planet is fixed at this orbital radius for its entire evolution. We discuss this assumption further in section \ref{subsec:HZ}. Figure \ref{fig:HZDist} shows the midpoint of the HZ calculated using equations (\ref{eq:HZDistance}) and (\ref{eq:EffFlux}) for an Earth mass planet. Both the luminosity and the effective temperature of a star decrease with increasing metallicity for a given stellar mass (see fig. \ref{fig:StructModels}). The net result is that the HZ midpoint gets closer to the host star with increasing metallicity for a given stellar mass. This is consistent with the findings of previous studies \citep{Danchi2013,Gallet2017}.

\begin{figure*}
	\includegraphics[trim=5mm 10mm 5mm 5mm,width=\textwidth]{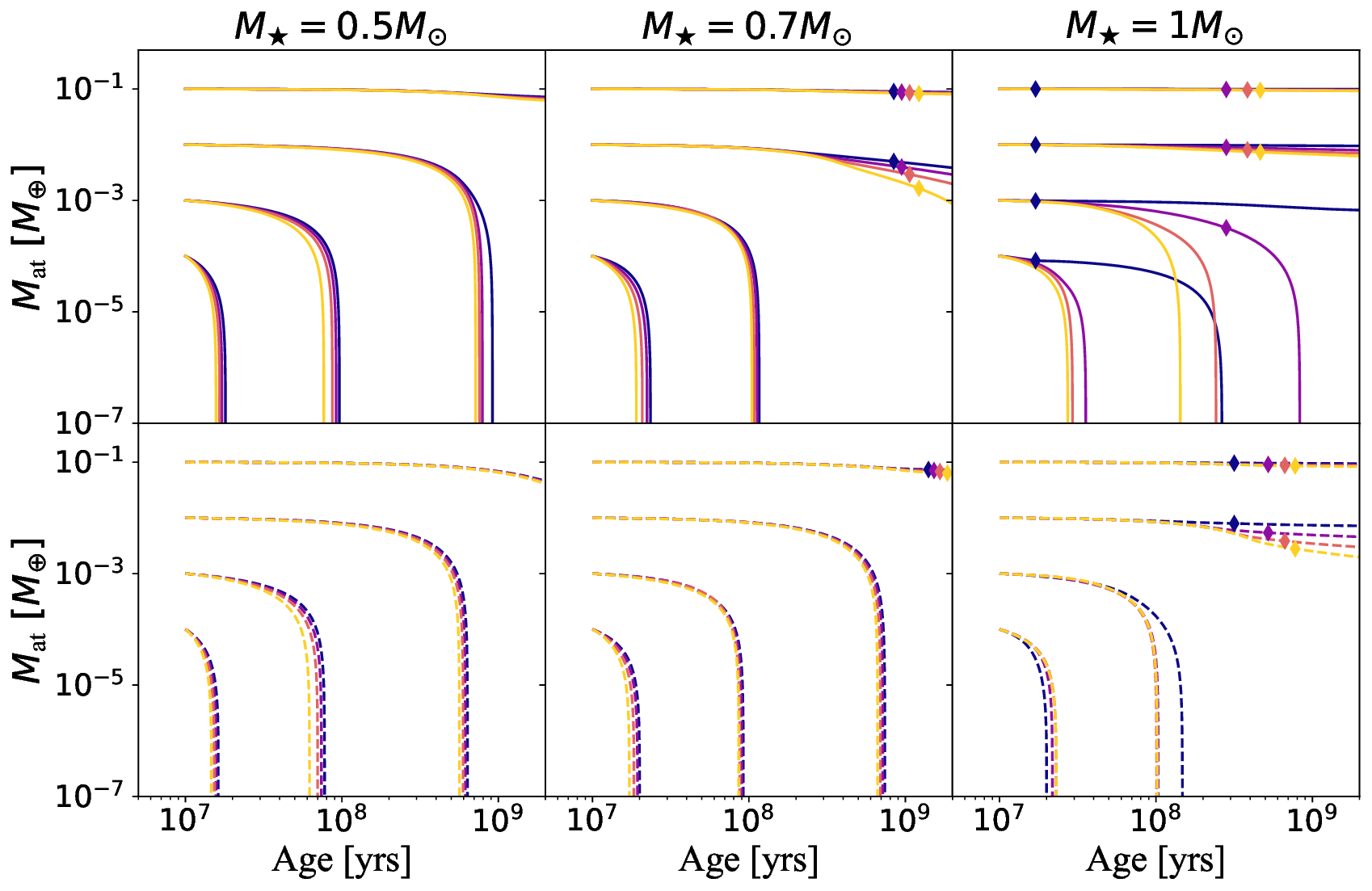}
    \caption{The atmospheric evolution of an Earth mass planet in the habitable zone of a 0.5$M_\odot$ (left), 0.7$M_\odot$ (middle) and 1$M_\odot$ star (right). We consider initial atmospheric masses of $f_{\rm at}=10^{-1}, 10^{-2}, 10^{-3}$ and $10^{-4}$. A range of metallicities and initial rotation periods are considered for the host star and are indicated by the same colour and line style scheme as fig. \ref{fig:RotEvo}. We separate the evolution for stars with initial rotation periods of 10 days (top row) and of 1 day (bottom row) for clarity.}
    \label{fig:EarthMassPlanet}
\end{figure*}

The bottom row of fig. \ref{fig:XUVLumEvo} shows the XUV flux evolution at the HZ midpoint, $F_{\rm XUV}(r_{\rm HZ,mid})$ for an Earth mass planet. There are two main features worth noting. First, $F_{\rm XUV}(r_{\rm HZ,mid})$ is lower for higher mass stars on average despite the fact that higher mass stars tend to have larger XUV luminosities (see row 3 of fig. \ref{fig:XUVLumEvo}). This is because higher mass stars have larger HZ distances. Second, $F_{\rm XUV}(r_{\rm HZ,mid})$ is generally larger for metal-rich stars than metal-poor stars, for a given stellar mass, initial rotation period and age, despite the fact that the same is not necessarily true of the XUV luminosity (see row 3 of fig. \ref{fig:XUVLumEvo}). Again, this can be attributed to the influence of the HZ distance since more metal-rich stars have closer in HZs than metal-poor ones.

\section{Photoevaporation of exoplanet atmospheres}
\label{sec:Exo}
In this section, we calculate how rapidly exoplanet atmospheres are evaporated by XUV radiation. When estimating evaporation rates, it is common to use an energy limited escape approach \citep[e.g.][]{Lammer2003,Selsis2007b}. However, in this work, we use the more physically motivated prescription of \citet{Johnstone2015}. Using hydrodynamic simulations, these authors constructed a scaling law for the rate at which a hydrogen atmosphere would be evaporated by XUV radiation which is given by

\begin{equation}
    \dot{M}_{\rm at} = am_{\rm H} M_{\rm pl}^b z_0^c (\log F_{\rm XUV})^{g(M_{\rm pl}, z_0)},
    \label{eq:MDotAt}
\end{equation}
where

\begin{equation}
    g(M_{\rm pl},z_0) = d M_{\rm pl}^ez_0^f.
    \label{eq:gFunc}
\end{equation}
Here, $m_{\rm H}$ is the hydrogen mass, $M_{\rm pl}$ is the planet mass and the constants are given by $a=1.858 \times 10^{31}$, $b=-1.526$, $c=0.464$, $d=4.093$, $e=0.249$ and $f=-0.022$. Lastly, $z_0=R_0-R_{\rm core}$, is the altitude at the base of the hydrodynamic simulation (see \citet{Johnstone2015} for further details) and can be calculated from 

\begin{equation}
    \log\left(\frac{R_0}{R_{\rm core}}\right) = (2.5f_{\rm at}^{0.4}+0.1)\left(\frac{M_{\rm pl}}{M_\oplus}\right)^{-0.7},
    \label{eq:R0Rcore}
\end{equation}
where $f_{\rm at}=M_{\rm at}/M_{\rm pl}$ is the ratio of the atmospheric mass to the planetary mass and $R_{
\rm core}$ is the core radius which is calculated assuming that the planet has the same density as Earth, i.e. $R_{\rm core} = (M_{\rm pl}/M_\oplus)^{1/3} R_\oplus$.

\begin{figure*}
	\includegraphics[trim=0mm 5mm 0mm 0mm,width=\textwidth]{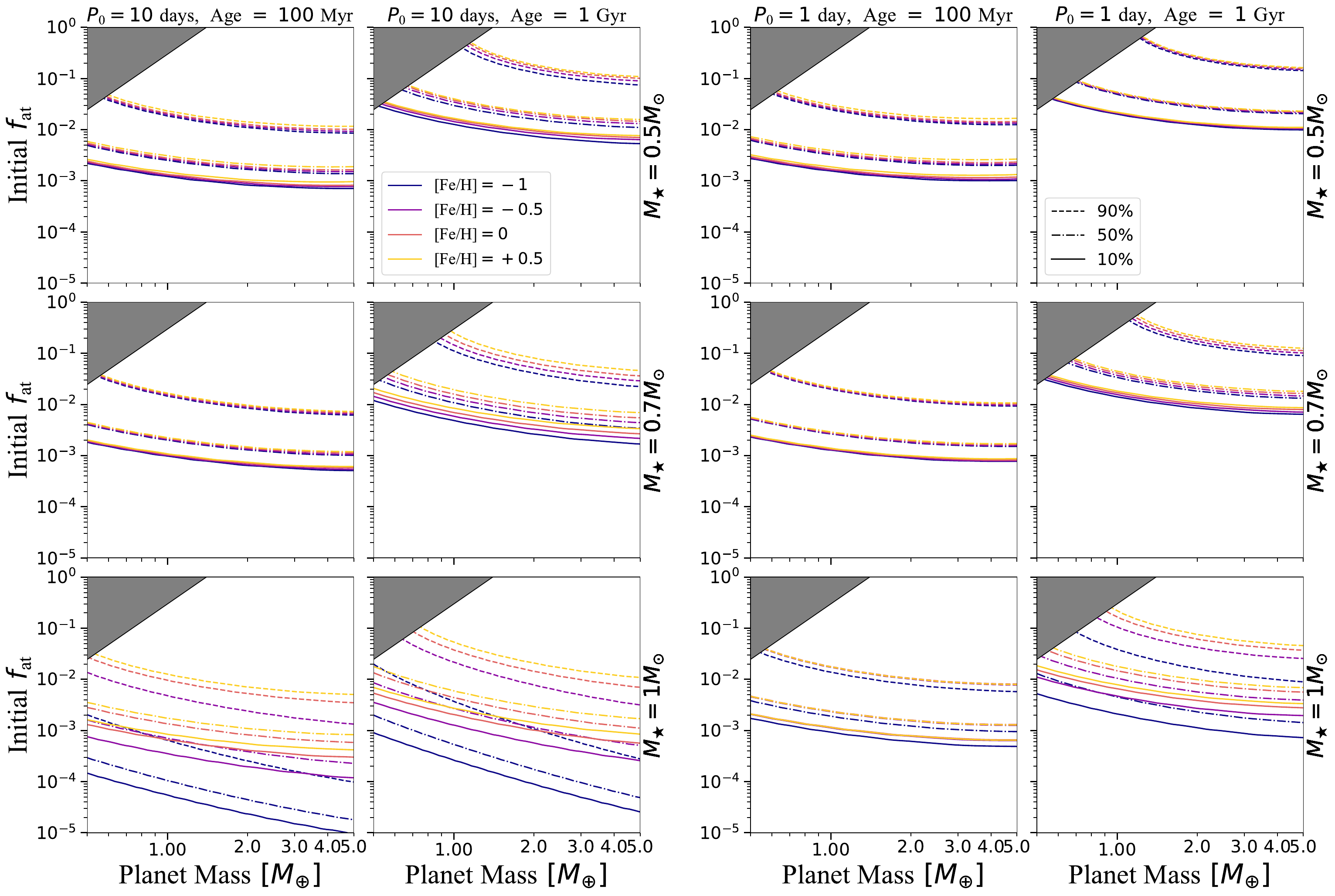}
    \caption{Contour plots showing the percentage of the initial atmosphere remaining for a range of initial atmospheric mass fractions and planetary masses as well as stellar masses, initial rotation periods and metallicities. Dashed, dot-dashed and solid lines indicate 90\%, 50\% and 10\% of the atmosphere remaining respectively. The line colours indicate the the metallicity of the host star and have the same meaning as fig. \ref{fig:RotEvo}. The top, middle and bottom rows represent planets in the habitable zones of 0.5$M_\odot$, 0.7$M_\odot$ and 1$M_\odot$ stars respectively. The first and second columns show the percentage of atmosphere remaining on planets orbiting stars with an initial rotation period of 10 days at 100 Myr and 1 Gyr respectively. The third and fourth columns show the percentage of atmosphere remaining on planets orbiting stars with an initial rotation period of 1 day at 100 Myr and 1 Gyr respectively. The grey shaded regions indicate initial atmospheric mass fractions larger than the maximum allowed atmospheric mass fraction as estimated by \citet{Johnstone2015}.}
    \label{fig:RangeOfPlanetMasses}
\end{figure*}

Before presenting the results of the atmospheric mass-loss modeling, it is worth briefly talking about the limitations of equation (\ref{eq:MDotAt}). At very low XUV fluxes, the planetary wind at the exobase becomes slower than the local escape velocity and equation (\ref{eq:MDotAt}) may become an unreliable predictor of the true atmospheric mass-loss rate. Based on the middle left panel of figure 1 in \citet{Johnstone2015}, this occurs somewhere between $F_{\rm XUV}=10\ {\rm erg\ s^{-1}\ cm^{-2}}$ and $50\ {\rm erg\ s^{-1}\ cm^{-2}}$. Since the lowest XUV fluxes occur at late ages, \citet{Johnstone2015} stop their atmospheric evolution modeling at 1 Gyr to avoid this problem and we also adopt this approach. However, even when stopping at 1 Gyr, there are times when our XUV fluxes drop into the $10\ {\rm erg\ s^{-1}\ cm^{-2}}$ to $50\ {\rm erg\ s^{-1}\ cm^{-2}}$ range or lower, most notably for the 1 $M_\odot$, [Fe/H]=-1, $P_{\rm 0}=$10 days case (see the bottom row of fig. \ref{fig:XUVLumEvo}). It is worth noting that in these few cases, the modeling may have an additional level of uncertainty, and equation (\ref{eq:MDotAt}) likely only represents an upper limit to the XUV driven atmospheric mass-loss rate. 

In fig. \ref{fig:EarthMassPlanet}, we consider the atmospheric evolution of an Earth mass planet with initial atmospheric mass fractions of, $f_{\rm at}=10^{-1}$, $10^{-2}$, $10^{-3}$ and $10^{-4}$, in the HZs of 0.5$M_\odot$, 0.7$M_\odot$ and 1$M_\odot$ stars where the XUV flux received by the planet is given by the results of section \ref{subsec:XUVFluxEvoResults}. To begin, we consider the atmospheric evolution of Earth mass planets around solar mass stars of solar metallicity, i.e. the orange, solar metallicity, tracks in the right hand panels. The evolution of Earth mass planets around solar mass and metallicity stars has previously been studied by \citet{Johnstone2015}. Our results in fig. \ref{fig:EarthMassPlanet} are qualitatively similar to the results of these authors. Planets with large initial atmospheric mass fractions of $f_{\rm at}=10^{-1}$ or $10^{-2}$ are able to hold onto nearly all of their atmosphere up to the modelled age of 1 Gyr. However, planets with lower initial atmospheric mass fractions can lose their entire atmospheres. In our model, this occurs on ${\sim}10^8$ year and ${\sim}10^7$ year time-scales for $f_{\rm at}=10^{-3}$ and $10^{-4}$ respectively. We also reconfirm the result of \citet{Johnstone2015} that the initial rotation period of the host star can have a significant impact on the atmospheric evolution of a planet. This is most evident when looking at the atmospheric evolution of a planet with an initial atmospheric mass fraction of $f_{\rm at}=10^{-3}$. Such a planet loses its entire atmosphere in $\sim 10^8$ years when orbiting a star with an initial rotation period of $P_0=1$ day but takes over twice as long to lose its atmosphere, $\sim 2.4\times 10^8$ years, when orbiting a star with an initial rotation period of $P_0=10$ days.

Next, we look at the impact of the stellar metallicity while still focusing on solar mass stars. Figure \ref{fig:EarthMassPlanet} shows that the atmospheres of planets around more metal-rich stars are generally evaporated more rapidly than those around metal-poor stars for a given initial atmospheric mass fraction and a given initial stellar rotation period. This is unsurprising given that we already demonstrated that the XUV flux at the HZ midpoint is generally greater for metal-rich stars than metal-poor stars in section \ref{subsec:XUVFluxEvoResults}. It is worth noting that, in some cases, the stellar metallicity has a larger impact on the atmospheric evolution than the initial rotation period of the host star. This is perhaps most evident for the $f_{\rm at}=10^{-3}$ case.

Lastly, we look at the impact of the host star mass on planetary atmospheric evolution. The left and middle panels of fig. \ref{fig:EarthMassPlanet} show the atmospheric evolution of planets in the HZs of  0.5$M_\star$ and 0.7 $M_\star$ stars respectively. In general, the atmospheres of planets in the HZs of 0.5$M_\star$ and 0.7 $M_\star$ stars are evaporated more rapidly than the atmospheres of planets in the HZs of solar mass stars due to the larger XUV flux these planets receive (see section \ref{subsec:XUVFluxEvoResults}). The trends discussed for planets around solar mass stars also hold for planets around lower mass stars. Specifically, planets lose their atmospheres more rapidly around stars that are metal-rich and with shorter initial rotation periods. We note, however, that the atmospheric evolution is less sensitive to the host star metallicity for planets around lower mass host stars than higher mass host stars. This is due to the sensitivity of the stellar convective properties to changes in metallicity (see discussion of this effect in section \ref{subsec:RotEvoResults}).

In fig. \ref{fig:RangeOfPlanetMasses}, we show contour plots showing the percentage of the initial atmosphere remaining for a range of planetary masses, initial atmospheric mass fractions, host star mass, host star initial rotation period and host star metallicity. When calculating the habitable zone midpoints for non-Earth mass planets, we linearly interpolated between the coefficients given in table \ref{tab:HZParams}. These plots are similar to the ones shown by \citet[][see their figure 5]{Johnstone2015}. The trends we identified for Earth mass planets (fig. \ref{fig:EarthMassPlanet}) all still hold when considering a range of planetary masses. Specifically, the planets that lose their atmospheres most rapidly are those with small initial atmospheric mass fractions, in the HZs of lower mass, metal-rich stars with rapid initial rotation. In addition to these trends, fig. \ref{fig:RangeOfPlanetMasses} shows that lower mass planets lose atmosphere more rapidly than higher mass planets which is in agreement with the findings of \citet{Johnstone2015}. Lastly, and similarly to before, we see that atmospheric evolution for all planetary masses is more sensitive to the stellar metallicity when the host star is more massive.

\section{Further Discussion}
\label{sec:Discussion}

\subsection{Habitable zone location considerations}
\label{subsec:HZ}
In this section, we discuss some of the caveats associated with our HZ location calculation. In section \ref{subsec:XUVFluxEvoResults}, we calculated the midpoint of the HZ at an age of 500 Myr for a range of stellar masses and metallicities. We then made the simplifying assumption that the exoplanet is located at this position for the entire lifetime of the planetary system in the rest of our modelling. However, in reality, the location of the HZ evolves over time as the physical properties of the host star evolves. A more sophisticated treatment of this problem would consider the concept of the continuously HZ. This is the orbital radius around a host star that stays within the HZ for some extended, or maximal, amount of time \citep[e.g.][]{Valle2014,Truitt2015,Gallet2017}. It is worth re-iterating that we calculated the HZ midpoint at 500 Myr because this age corresponds to the main sequence evolutionary phase for the stars we consider in this work. As such, we would expect it to be representative of the HZ location throughout the majority of the, multi-billion year, main sequence lifetimes of the stars we consider since stars do not evolve significantly during this period. Although, we have not explicitly determined the continuously HZ location, i.e. the planetary orbital radius that maximises the time spent in the HZ, for each of our models, it is likely to be close to the HZ midpoint at 500 Myr for this reason. One nuance this does not account for is when the continuously HZ begins and how long it exists for. The work of \citet{Gallet2017} suggests that the continuously HZ comes into existence later for more metal-rich and lower mass host stars. Therefore, it is possible that, when we perform our photoevaporation calculations, the planets around these types of stars do not begin within the continuously HZ. However, exploring the effects of this scenario is beyond the scope of our current work.

Numerous prescriptions exist in the literature to determine the location of the HZ \citep[e.g.][]{Underwood2003,Selsis2007a,Kopparapu2014}. As such, it is worth considering whether our choice to use the prescription of \citet{Kopparapu2014} could drastically affect our results. \citet{Gallet2017} tested a number of different prescriptions and found that the predicted HZ locations only changed by ${\sim}10\%$, at least for solar mass stars. Such small differences between the prescriptions gives us confidence that our particular choice of prescription is unlikely to affect overall conclusions of our work.

\subsection{Convective turnover times}
\label{subsec:TurnoverTimes}
Convective turnover times are notoriously difficult to estimate \citep[e.g.][]{Corsaro2021,Gossage2025} since they cannot be directly measured. In our work, we require turnover time estimates for two purposes. The first is to calculate the spin-down torque in our rotation evolution modeling (see equation (\ref{eq:MattTorque})) and the second is to re-calibrate the Rossby numbers for our X-ray versus Rossby number relation (section \ref{subsec:DetermineXUV}). Many different prescriptions for estimating turnover times exist in the literature. In both cases, we use the prescription of \citet{Cranmer2011}, where the turnover time is parameterised as a function of stellar effective temperature (see equation (\ref{eq:Turnover})). We choose this prescription because it was also used by \citet{Matt2015}, since our rotation evolution model is heavily based on theirs.

In principle, we could have directly calculated turnover times from the structure models of \citet{Amard2019} in our rotation evolution modeling. This would have the benefit of making the turnover time estimates more self-consistent with the rest of the rotation modeling since the other physical stellar parameters are also obtained from the \citet{Amard2019} structure models. The drawback is that the X-ray data set we use from \citet{Wright2011} does not contain all the information we require in order to estimate turnover times using the \citet{Amard2019} structure models. Therefore, we use the prescription of \citet{Cranmer2011} for both purposes so that the turnover times and, therefore, the Rossby numbers used in our X-ray versus Rossby number relation are consistent with the ones in our rotation evolution modeling. This is important in order to properly model the XUV evolution (section \ref{sec:ActEvo}).

The \citet{Cranmer2011} prescription is constructed using solar metallicity structure models of \citet{Gunn1998}. Therefore, one could question whether it properly encapsulates the dependence that convective turnover times have on the metallicity of the star. This point is important to consider given that the central theme of this work is the metallicity dependent behaviour of stellar and planetary evolution. Two facts give us confidence in our choice to use the \citet{Cranmer2011} prescription. Firstly, the convective turnover times of main sequence stars are largely, though not entirely, a function of effective temperature, even across stars of different metallicities \citep{Amard2020RotEvo}. As such, a prescription that is parameterised in terms of the effective temperature likely captures most of the metallicity dependence that turnover times have. Indeed, we previously explored the metallicity dependence of the \citet{Cranmer2011} prescription in the Appendix of \citet{See2021}. Figure 9 of \citet{See2021} shows that the \citet{Cranmer2011} turnover times do indeed follow the same metallicity and stellar mass trends as turnover times from the models of \citet{Amard2019} in the range of stellar masses we have explored in this work, albeit with an approximately constant offset between the two. This offset would not impact the rotation evolution modeling in this work since the turnover times are normalised by the solar value. Secondly, as discussed in section \ref{subsec:RotEvoResults}, our rotation evolution modeling is in broad agreement with the results of \citet{Amard2020RotEvo} who used self-consistently calculated turnover times from their structure models. This also suggests that the \citet{Cranmer2011} prescription is doing a reasonable job of capturing the metallicity dependence of convective turnover times, at least in the range of stellar masses we consider in this work.

\subsection{Link between stellar composition and planetary properties}
\label{subsec:StarPlanetLink}
 Within a planetary system, all the objects are formed from the same original material. Therefore, one might expect that, within a given system, the elemental abundances of stars, planets and even other objects, e.g. meteorites, should be similar, especially for refractory elements. Indeed, such a link has been demonstrated for the Sun, Earth and solar system meteorites \citep{Lineweaver2009,Asplund2009} and evidence suggests that it is also the case for extra-solar systems \citep{Bonsor2021,Schulze2021}. This link is often used to provide further constraints on the interior structures of exoplanets \citep[e.g.][]{Dorn2015,Santos2015,Hinkel2018,Putirka2019,Otegi2020,Wang2022,Spaargaren2023,Unterborn2023}.

In this work, we have not attempted to link the properties of our modelled exoplanets to the properties of their host stars. Rather, we simply explored a range of planetary masses and assumed that the planetary cores all had the same density as Earth's (see section \ref{sec:Exo}). However, there is evidence to suggest that a host star's composition may play a role in determining both a planet's mass \citep{Nielsen2023} and its density \citep{Adibekyan2021,Liu2023}. Simultaneously, it is also worth noting that expanded exoplanet sample sizes may yet indicate that the link between planetary and host-star properties is not as strong as previously thought (Turner et al. in prep). While it is outside the scope of the current investigation, future work could explore the impact that linking planetary properties to the host star composition could have on atmospheric loss rates, if such a link proves to be robust. For instance, scaling the planetary density as a function of stellar composition would affect the planetary core radius, and ultimately, the atmospheric mass-loss rate (see equations (\ref{eq:MDotAt})-(\ref{eq:R0Rcore})).

\section{Conclusions}
\label{Sec:Conclusions}
In recent years, a number of studies have begun to reveal how the metallicity of low-mass stars affects their rotation and activity. The picture that has emerged is that more metal-rich stars are more magnetically active \citep{See2021,See2023} due to the deeper convection zones they are expected to have. Theoretical modeling \citep{Amard2020RotEvo} and ensemble studies of stars in the Kepler field \citep{Amard2020Kepler,See2024} suggest that this metallicity dependence of activity should also impact the rotation evolution in low-mass stars resulting in more metal-rich stars spinning more slowly at late ages.

In this work, we build on these previous studies to investigate how stellar metallicity affects the evolution of stellar rotation, activity and exoplanetary atmospheres using parametric evolution models. Our main conclusions are as follows:

\begin{enumerate}
    \item For a given mass, more metal-rich stars spin down to slower rotation periods at late ages which is in agreement with the results of \citet{Amard2020RotEvo}.
    \item For a given initial rotation period and mass, the stellar X-ray to bolometric luminosity ratio, $R_{\rm X} = L_{\rm X}/L_{\rm bol}$, is larger for more metal-rich stars at all phases of stellar evolution.
    \item In contrast, there are periods of time when the stellar X-ray luminosity, $L_{\rm X}$, and the XUV luminosity, $L_{\rm XUV}$ are larger in metal-poor stars. This is due to the fact that metal-poor stars have larger bolometric luminosities than metal-rich stars.
    \item For a given initial rotation period, mass and age, the XUV flux at the midpoint of the habitable zone, $F_{\rm XUV}(r_{\rm HZ,mid})$, is nearly always larger for more metal-rich stars. This is due to the habitable zone being closer to the host star for more metal-rich stars.
    \item The rate at which exoplanet atmospheres are evaporated depends on many parameters including the mass, initial rotation period and metallicity of the host star as well as the mass and initial atmospheric mass fraction of the planet. For exoplanets in the habitable zones of their host star, the rate of evaporation is nearly always larger for more metal-rich stars for a given host star mass, host star initial rotation period, planetary mass and planetary initial atmospheric mass fraction.
    \item The evolution of stellar rotation and activity as well as exoplanetary atmospheres are most sensitive to changes in stellar metallicity in host stars with high masses. This is because high mass stars have thinner convective zones and so even a small change in the metallicity can result in large fractional change in the convective properties of the star.
\end{enumerate}

\section*{Acknowledgements}
We thank the anonymous referee for useful comments that helped improve the quality of this work. We also thank T. G. Wilson and C. P. Johnstone for useful discussions that helped improved this manuscript. V. S. and O. H. acknowledges support from the European Space Agency (ESA) as ESA Research Fellows. V. S. also acknowledges support from the European Research Council (ERC) under the European Union’s Horizon 2020 research and innovation programme (CartographY G.A. n. 804752). C.F. is funded by the University of Bristol School of Physics PhD Scholarship Fund and acknowledges support from ESA and the University of Leiden through the LEAPS program. L. A. acknowledges support from the Centre National des Etudes Spatialees (CNES) through the PLATO/AIM grant, the Swiss National Science Foundation (SNF; Project 200021L-231331) and the French Agence Nationale de la Recherche (ANR-24-CE93-0009-01) ``PRIMA - PRobing the origIns of the Milky WAy’s oldest stars''. 

\emph{Software:} \texttt{matplotlib} \citep{Hunter2007}, \texttt{numpy} \citep{Harris2020}, \texttt{scipy} \citep{Virtanen2020}

\section*{Data Availability}

This work has not generated any new datasets.



\bibliographystyle{mnras}
\bibliography{MetActRotExo} 





\bsp	
\label{lastpage}
\end{document}